\documentclass[aps,twocolumn,superscriptaddress,floatfix,prl]{revtex4-1}

\usepackage{lipsum}
\usepackage{graphicx}
\usepackage{amsmath,amssymb}
\usepackage{braket}
\usepackage{mathtools}
\usepackage{hyperref}
\usepackage{multirow}
\usepackage{color}
\usepackage[normalem]{ulem}
\usepackage{amsfonts}
\usepackage{float}
\usepackage{pdfpages} 
\makeatletter
\usepackage{subfigure}

\def\beq{\begin{equation}}
\def\eeq{\end{equation}}
\def\bea{\begin{eqnarray}}
\def\eea{\end{eqnarray}}

\AtBeginDocument{\let\LS@rot\@undefined}
\makeatother

\begin{document}

\title{Subdiffusive hydrodynamics of nearly-integrable anisotropic spin chains}

\author{Jacopo De Nardis}
\affiliation{Laboratoire de Physique Th\'eorique et Mod\'elisation, CNRS UMR 8089,CY Cergy Paris
Universit\'e, 95302 Cergy-Pontoise Cedex, France}

\author{Sarang Gopalakrishnan}
\affiliation{Department of Physics, The Pennsylvania State University, University Park, PA 16802, USA}

\author{Romain Vasseur}
\affiliation{Department of Physics, University of Massachusetts, Amherst, MA 01003, USA}

\author{Brayden Ware}

\affiliation{Joint Quantum Institute, NIST/University of Maryland, College Park, Maryland 20742, USA}
\affiliation{Joint Center for Quantum Information and Computer Science,
NIST/University of Maryland, College Park, Maryland 20742, USA}

\begin{abstract}

We address spin transport in the easy-axis Heisenberg spin chain subject to integrability-breaking perturbations. We find that spin transport is subdiffusive with dynamical exponent $z=4$ up to a timescale that is parametrically long in the anisotropy. In the limit of infinite anisotropy, transport is subdiffusive at all times; for large finite anisotropy, one eventually recovers diffusion at late times, but with a diffusion constant \emph{independent} of the strength of the integrability breaking perturbation. We provide numerical evidence for these findings, and explain them by adapting the generalized hydrodynamics framework to nearly integrable dynamics. Our results show that the diffusion constant of near-integrable interacting spin chains is  generically not perturbative in the integrability breaking strength.

\end{abstract}
\vspace{1cm}

\maketitle

Many strongly interacting one-dimensional and quasi-one-dimensional experimental systems are approximately described by integrable models such as the Heisenberg and Hubbard models~\cite{Wu2016,PhysRevLett.111.137205,Scheie2021,Mourigal2013,Salomon2018,Wu2019}. Although integrable systems are in some sense exactly solvable, the problem of characterizing their long-distance, late-time hydrodynamic response at nonzero temperature remained largely open until the recent advent of generalized hydrodynamics (GHD)~\cite{Doyon, Fagotti, SciPostPhys.2.2.014, PhysRevLett.119.020602,  BBH0, BBH,PhysRevLett.119.020602, GHDII, doyon2017dynamics, solitongases,PhysRevLett.119.195301,2016arXiv160408434Z, PhysRevB.96.081118,PhysRevB.97.081111, PhysRevLett.120.164101, dbd1, ghkv, DeNardis_SciPost, GV19,Balasz, horvath2019euler, PhysRevB.100.035108,2019arXiv190601654B,10.21468/SciPostPhys.8.3.041,ruggiero2019quantum,PhysRevLett.125.070602, 2104.04462,PhysRevLett.122.090601,PhysRevLett.126.090602,2020arXiv200906651M}, as well as modern numerical methods~\cite{Bertini2021}. It was found, remarkably, that although integrable systems possess stable, ballistically propagating quasiparticles, certain quantities (such as spin in the Heisenberg model) are transported diffusively or superdiffusively~\cite{PhysRevLett.106.220601,lzp,PhysRevLett.122.210602,idmp,GV19,PhysRevLett.123.186601,gvw,vir2019,dmki,2019arXiv190905263A,dupont_moore,Bulchandani2021,PhysRevB.102.115121,PhysRevLett.125.070601,Ilievski2021,PhysRevLett.127.057201}. Moreover, there are dynamical phase transitions between the ballistic and diffusive transport ``phases.'' The GHD framework has been extended to incorporate some of these transport anomalies. In addition, the prediction of anomalous spin transport has recently been experimentally verified in quantum magnets~\cite{Scheie2021} and ultracold atomic gases~\cite{2107.00038}.

Given that experiments are never described by perfectly integrable systems, it is natural to ask how weak integrability-breaking perturbations affect these anomalies. Incorporating such perturbations into GHD is in general a challenging open problem, despite much recent progress~\cite{Friedman2020,2020arXiv200411030D,Bastianello_noise,Bastianello2020,2103.11997}. Qualitatively, integrability-breaking perturbations endow the ballistic quasiparticles with a finite lifetime, after which they scatter or decay. For quantities such as energy that are transported ballistically in the integrable limit, integrability-breaking generically renders transport diffusive: the zero-frequency singularity or ``Drude peak'' associated with ballistic transport broadens into a Lorentzian feature of width set by the strength of the integrability-breaking perturbation, or alternatively by the life-time of the quasiparticles~\cite{PhysRevLett.115.180601,huangkarrasch,PhysRevB.94.245117,2103.11997,Friedman2020,2020arXiv200411030D,Bastianello2020,PhysRevLett.127.057201,tang2018,PhysRevX.9.021027,PhysRevB.103.L060302,PhysRevLett.125.180605,2020arXiv201207849L,2021arXiv210908390Z}, in full analogy with standard quantum Boltzmann equation~\cite{landau1981course}.

In the present work we address what happens to quantities that are \emph{already diffusive} in the  interacting-integrable limit. As we shall see the mechanism at play in this case is different from the usual broadening of Drude weight:  transport coefficients are discontinuous functions of the integrability breaking coupling. 

As a concrete example, we consider spin transport in the anisotropic Heisenberg (or XXZ) spin chain, governed by the Hamiltonian
\beq\label{hxxz}
H_{\mathrm{XXZ}} = J \sum_i (S^x_i S^x_{i+1} + S^y_i S^y_{i+1} + \Delta S^z_i S^z_{i+1}),
\eeq
where $S^\alpha_i = \sigma^\alpha_i/2$ represents the $\alpha = (x,y,z)$ spin-$1/2$ operator on site $i$, $J$ is an overall energy scale (set to 1 in what follows), and $\Delta$ is the anisotropy parameter. At nonzero temperature, the equilibrium state is always a paramagnet, and the late-time dynamics is qualitatively the same regardless of the sign of $J$. 

\begin{figure}
    \centering
   \includegraphics[width = 0.95\columnwidth]{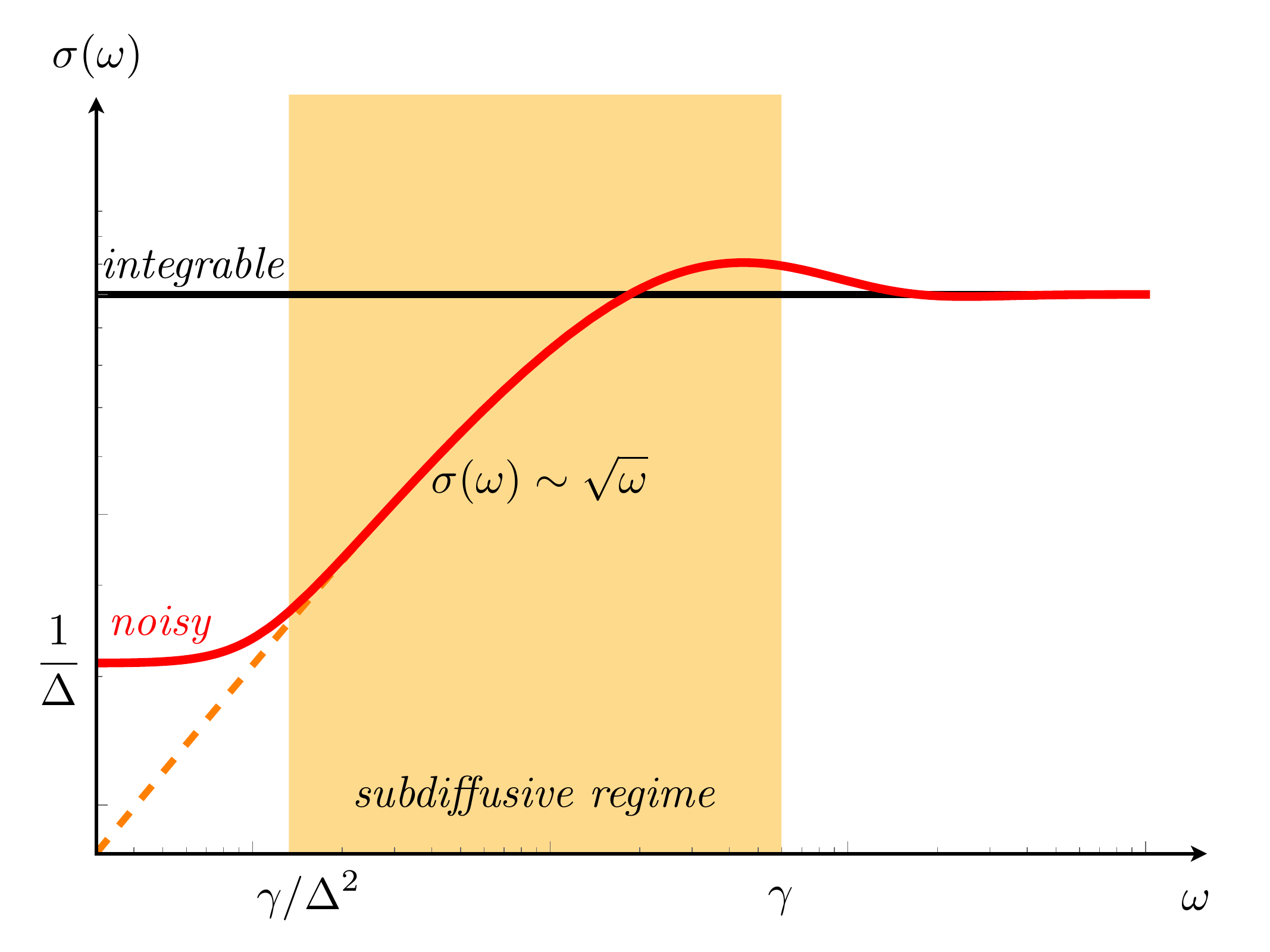}
    \caption{{\bf Anomalous low-frequency spin conductivity in the noisy XXZ chain} (in log-log scale). In the frequency regime $\frac{\gamma}{\Delta^2} \ll \omega \ll \gamma$, spin transport is subdiffusive with $\sigma(\omega) \sim \sqrt{\omega}$, corresponding to dynamical exponent $z=4$. At very low frequency, the conductivity eventually saturates to a finite d.c. value that is {\em independent} of the noise strength $\gamma$ (see also additional plot in \cite{suppmat}).  }
    \label{fig:cartoon}
\end{figure}

When $\Delta > 1$, transport is diffusive in the purely integrable limit, with a diffusion constant $D_{\mathrm{XXZ}}(\Delta)$ that is exactly known \cite{DeNardis_SciPost,GV19}; as $\Delta \to \infty$, $D_{\mathrm{XXZ}}(\Delta)$ approaches a nonzero constant when rescaling the temperature so as to keep $\beta \Delta$ fixed.
We address here what happens when integrability is weakly broken, with some generic perturbation of strength $\gamma$. Explicitly, we consider the effect of local spin dephasing, described by the Hamiltonian $H = H_{\mathrm{XXZ}} + \sqrt{\gamma} \sum_i \eta_i(t) S^z_i$, with white noise $\eta$. Since the integrable and the perturbed non-integrable model are both diffusive, one might expect that the diffusion constant $D_{\mathrm{XXZ}}(\gamma,\Delta)$  simply picks up $\gamma$-dependent corrections $D_{\mathrm{XXZ}}(\gamma,\Delta) = D_{\mathrm{XXZ}}(\Delta) + O (\gamma^r)$, with some power $r$. What we find is much more striking: for arbitrary $\gamma > 0$, the true diffusion constant $D(\gamma, \Delta)$ is strongly suppressed at large $\Delta$, and approaches zero as $\Delta \to \infty$. Moreover, $D(\gamma, \Delta)$ does not depend on the strength of the integrability-breaking perturbation. In the limit $\Delta \to \infty$, the dynamics is \emph{subdiffusive}, with the spacetime scaling $x \sim t^{1/4}$. For large finite $\Delta$, subdiffusion occurs over a timescale that grows with $\Delta$, before crossing over to diffusion at the latest times.  In effect, the integrability-breaking perturbation moves spectral weight from the spin conductivity from very low frequencies to a peak at frequency $\sim \gamma$, as shown in Fig.~\ref{fig:cartoon}.

\emph{Observable}.---We consider linear-response transport under the Hamiltonian~\eqref{hxxz}. We expect that our results hold at any nonzero temperature, but for simplicity we will work in the high-temperature limit, where the frequency-dependent conductivity is simply related to the autocorrelation function of the spin current operator $\hat J(t) = \sum_x \hat j_x(t)$:
\beq
T\sigma(\omega) = \int_0^\infty dt \, \langle \hat{J}(t) \hat{j}_0(0)\rangle e^{i \omega t}.
\eeq
Under diffusive dynamics (i.e., for Eq.~\eqref{hxxz} with no integrability-breaking perturbation), the current-current correlator decays on a finite timescale $\tau$; therefore the conductivity is close to its d.c. value, $\sigma(\omega) \simeq D \chi$ for $\omega \tau \ll 1$, with $\chi$ the static susceptibility. For finite $\gamma$, remarkably, the correlator \emph{overshoots}, so the late-time current, on timescales $\gamma t > 1$, becomes \emph{anticorrelated} with the early-time current. Thus the integrability-breaking perturbation shifts spectral weight from very low frequencies to a peak at $\omega \sim \gamma$ (Fig.~\ref{fig:cartoon}). 

\emph{Quasiparticle picture}.---We now explain the origin of this phenomenon, in terms of the quasiparticle structure of Eq.~\eqref{hxxz}. For $\Delta > 1$ this model has infinitely many quasiparticle species, called ``strings.'' These strings are easiest to visualize near the ferromagnetic vacuum, where they simply correspond to domains of various sizes. Under the integrable dynamics, the number of domains of each size is separately conserved. At large $\Delta$, a domain of size $s$ can only move collectively, via an $s^{\rm th}$ order process in perturbation theory, with an effective tunneling amplitude $\sim \Delta^{1-s}$. Because of integrability, even in a high-temperature thermal state, these strings remain stable, and their characteristic velocity scale does not change appreciably, although their other properties are highly renormalized, as we now discuss. 

\emph{$\Delta = \infty$ limit}.---It is instructive to consider the $\Delta = \infty$ limit first; this limit is sometimes called the ``folded XXZ model''~\cite{10.21468/SciPostPhysCore.4.2.010, zadnik2021folded, pozsgay2021integrable, bidzhiev2021macroscopic}. Here, the quasiparticle picture simplifies: all strings with $s > 1$ are frozen, and the only dynamics is due to the $s = 1$ strings, or magnons, moving in a static background of spin domains with velocity $v = O(1)$ [see also Ref.~\cite{PhysRevB.103.094303}]. In the integrable limit, magnons move ballistically. However, as a magnon moves through the system, the spin it carries fluctuates: e.g., when it is moving through a spin-up domain, it does so as a minority spin-down particle, but on passing into a spin-down domain it becomes a minority \emph{spin-up} particle. On average there are equally many up and down domains in a high-temperature state, so the magnon carries no net spin, hence the absence of ballistic transport. However, over a time $t$, the region it traverses (of size $|v t|$) has a net magnetization $1/\sqrt{|vt|}$ from equilibrium thermal fluctuations. Thus the effective spin carried by the magnon over this distance is $O(1/\sqrt{|vt|})$. Since, in time $t$, the magnon transports an amount of spin $\sim t^{-1/2}$ over a distance $\sim t$, spin transport is diffusive with an $O(1)$ diffusion constant. At infinite temperature, this diffusion constant has the closed-form expression $D(T = \infty, \Delta = \infty) = 4/(3\pi)$~\cite{DeNardis_SciPost,GV19}. 

We now consider, heuristically, what happens when integrability is broken by a generic local perturbation. In principle, the perturbation could either relax the momentum of a magnon, or change the number of magnons. As we will discuss below, the latter process becomes impossible for generic, sufficiently local perturbations at $\Delta = \infty$. Thus the only thing perturbations can do is scatter magnons. One can easily adapt the previous argument to the case of diffusive magnons: in a time $t$, they have moved by an amount $\sqrt{Dt}$, over which the net magnetization fluctuations are $(Dt)^{-1/4}$; thus, one would spin transport to be \emph{subdiffusive}, with the scaling $x \sim t^{1/4}$, corresponding to a conductivity scaling as:
\begin{equation}
\sigma(\omega) \sim \sqrt{\omega}.
\end{equation}
 As a function of time, transport would be diffusive until the mean free time of the magnon (which depends on the integrability-breaking parameter), and subdiffusive thereafter.

\emph{Mapping to constrained models}.---This heuristic argument can be put on a firmer footing if one notes that $H_{\mathrm{XXZ}}$ at $\Delta = \infty$ is a kinetically constrained model. In this model, the total number of domain walls is strictly conserved (since creating or destroying domain walls costs infinite energy), and the only allowed spin moves are those that respect this conservation law --- similar constraints have been considered recently in Refs.~\cite{PhysRevLett.124.207602,10.21468/SciPostPhysCore.4.2.010,PhysRevLett.125.245303,pozsgay2021integrable,2021arXiv210600696P,2021arXiv210804845B,2021arXiv210802205S}. We have checked that \emph{any} perturbations acting on four or fewer sites that conserve the domain wall number \emph{also} conserve the number of magnons. Thus, as we anticipated, the only thing an integrability-breaking perturbation (acting on fewer than five sites) can do is scatter magnons, supporting the heuristic argument above. 

Interestingly, one can go further, by considering dynamics that is constrained and conserves magnon number, but is otherwise random. This corresponds to the case where integrability is \emph{strongly} broken. The transport properties of this stochastic model were very recently computed by a Markov-matrix method in Ref.~\cite{2021arXiv210802205S}; the subdiffusive transport exponent $x \sim t^{1/4}$ was computed from the low-energy spectrum of the Markov matrix in that work. 
To check that the phenomenon we are considering is due specifically to \emph{magnon} physics (and not a generic consequence of the domain-wall conservation law, which holds until exponentially-long time scales $\sim {\rm e}^{\Delta}$~\cite{Abanin:2017aa}), we have explored random dynamics that obeys the domain-wall constraint but allows for moves on five or more sites. Such gates do not in general conserve magnon number: for example, they connect the configuration $\ldots\downarrow\downarrow\uparrow\uparrow\uparrow\downarrow\downarrow\uparrow\downarrow\ldots$ to $\ldots\downarrow\downarrow\uparrow\uparrow\downarrow\downarrow\uparrow\uparrow\downarrow\ldots$, turning two $2$-strings into a magnon and a $3$-string. We find that the random constrained dynamics is diffusive for all gate sizes $\geq 5$, highlighting the central role that magnon physics and proximity of integrability play in subdiffusion~\cite{suppmat}. In particular, we emphasize that the physics at play here is unrelated to other types of $z=4$ dynamics in the presence of fracton-like constraints recently discussed in the literature~\cite{PhysRevX.10.011042, PhysRevB.100.214301, PhysRevResearch.2.033124, PhysRevLett.125.245303}. 

\begin{figure}[tb]
\begin{center}
\includegraphics[width=0.95\columnwidth]{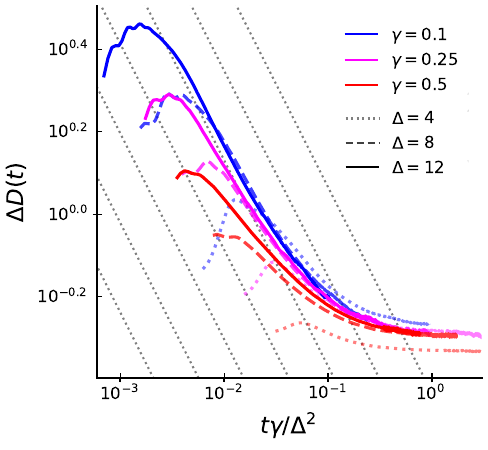}
\caption{{\bf Crossover of the time-dependent diffusion in the noisy XXZ spin chain.} We plot the diffusion constant $D(t)$ times $\Delta$ at infinite temperatures, for the time evolution in the presence of on-site noise of strength $\gamma \in \{0.1, 0.25, 0.5\}$ at large anisotropy $\Delta \in \{4, 8, 12\}$. The axes are rescaled to test the theoretical prediction~\eqref{eqCollapse}. For small $\gamma$ and large $\Delta$, we find good agreement: in particular, the diffusion constant saturates to a $\gamma$-independent value at long times, and is compatible with subdiffusion with $D(t) \sim t^{-1/2}$ (shown as dotted lines) for $\gamma^{-1} \ll t \ll \Delta^2/\gamma$.}
\label{noisefig}
\end{center}
\end{figure}

\begin{figure}[tb]
\begin{center}
\includegraphics[width=0.95\columnwidth]{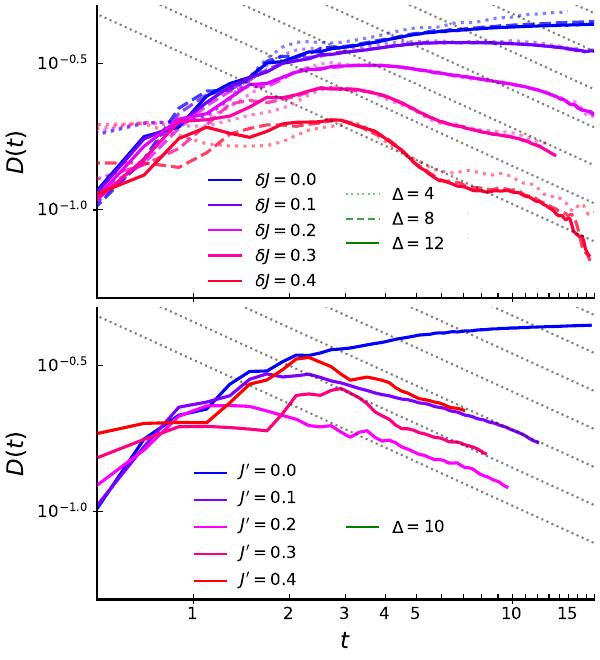}
\caption{{\bf Hamiltonian perturbations.} 
Time-dependent diffusion constant $D(t)$ with the couplings staggered by an amount $\delta J $ at large anisotropy $\Delta \in \{4, 8, 12\}$ (top)  and next-nearest-neighbour couplings $J'$ at $\Delta=10$ (bottom). Consistent with the theoretical prediction, $D(t)$ exhibits a subdiffusive regime consistent with $D(t) \sim t^{-1/2}$ (shown as dotted lines) at intermediate times whenever $\delta J >0$ or $J'>0$. 
}
\label{hamfig}
\end{center}
\end{figure}
\emph{Finite $\Delta$ and noise}.---
We now turn to finite $\Delta$. To discuss this case we need to specify the noise model more explicitly.
We first consider the simplest integrability-breaking perturbation, namely uncorrelated noise. We take $H = H_{\mathrm{XXZ}} + \sqrt{\gamma} \sum_i \eta_i(t) S^z_i$, where the noise $\eta$ has the properties $\langle \eta_i \rangle = 0$ and $\langle \eta_i(t) \eta_j(0)\rangle =  \delta(t) \delta_{ij}$. This noise backscatters magnons at a rate $\sim \gamma$. It can also create magnons out of strings, by the following process: one end of a larger string virtually hops away from the rest of the string by one site, with amplitude $1/\Delta$, and is put on shell by the noise, giving a transition rate $\gamma/\Delta^2$. 

The quasiparticle picture is modified in this case as follows. Magnons that were created at the initial time propagate ballistically until they hit their mean free path, contributing to diffusive spin transport. At later times they contribute only through subdiffusion. However, new magnons are created at a rate $\gamma/\Delta^2$, and then propagate for a time $1/\gamma$ before back-scattering; thus at any time some fraction of magnons are contributing to diffusive transport. In addition to magnons, one should also consider the contribution to transport due to mobile larger strings; however, except close to $\Delta = 1$, the velocity of these large strings is exponentially suppressed, and in any case they will also contribute subdiffusively to transport by exactly the same reasoning as we used for magnons.

The nature of the transport crossovers can be understood by a straightforward scaling argument. Let us define a time-dependent diffusion constant via the relation $D(t) =\frac{1}{2} \frac{d \delta x^2}{dt}$ with $\delta x(t)$ the variance of the spin structure factor ${\cal C}(x,t)=\langle \sigma_z(x,t) \sigma_z(0,0)\rangle$. If $\gamma$ is small enough, then on a timescale $1/\gamma$, $D(t) = D_{\mathrm{XXZ}}(\Delta)$ is some well-defined $O(1)$ number which is approximately independent of $\Delta$ for $\Delta \gg 1$. On timescales $1/\gamma \ll t \ll \Delta^2/\gamma$, the dynamics will be subdiffusive, so $D(t) \sim t^{-1/2}$. Enforcing continuity at $t_\gamma = 1/\gamma$, we find that $D(t) \sim (\gamma t)^{-1/2}$, independently of $\Delta$. Finally, enforcing continuity at the later crossover timescale $t_\star = \Delta^2/\gamma$, we get the asymptotic diffusion constant $D_\infty \sim 1/\Delta$, with no $\gamma$ dependence. We thus expect the scaling form (valid for $\gamma \ll1$, $\omega \ll \gamma$, $\Delta \gg 1$ and $ \frac{\omega \Delta^2}{\gamma}$ fixed) for the conductivity:
\begin{equation}
\sigma(\omega,\gamma,\Delta) = \frac{1}{\Delta} f\left( \frac{\omega \Delta^2}{\gamma}\right), \label{eqCollapse}
\end{equation}
with $f(\infty) = {\rm const.}$, and $f(x) \sim x^{1/2}$ for $x \ll 1$. Equivalently, we expect a time-dependent diffusion constant scaling as $D=\frac{1}{\Delta} g\left( \frac{t \gamma}{\Delta^2}\right)$. 
As shown in Fig.~\ref{noisefig}, direct numerical calculations of $D(t)$ using matrix product operator (MPO) methods~\cite{schollwoeck} collapse well onto this scaling form even for relatively large $\gamma$ and intermediate $\Delta$. Curiously, our numerical results suggest that $\lim_{y \to \infty} g\left( y \right) = \frac{1}{2}$, although we do not have a theoretical prediction for this value.  

\emph{Hamiltonian perturbations}.---So far, we included spin dephasing both to allow for well-converged numerical checks and to simplify the theoretical analysis by allowing us to neglect energy conservation. We now turn to generic Hamiltonian perturbations. For this class of perturbations, we can derive general lower bounds on the crossover timescale (or equivalently \emph{upper} bounds on the diffusion constant), but these might not be tight for all perturbations. The first lower bound, which is just a consequence of energy conservation, can be derived as follows. As we saw above, the physics that sets this crossover timescale is the creation or destruction of magnons. Consider the simplest such process, in which two $2$-strings collide to create a magnon and a $3$-string. This process must conserve energy; since the initial and final states have the same number of domain walls, it suffices to consider the kinetic energy of the magnons. The bandwidth of the $2$-string is suppressed by a factor $1/\Delta$ (at large $\Delta$) relative to that of the magnon. (The bandwidth of the $3$-string is suppressed by yet another factor of $\Delta$ and is negligible.) Thus, conservation of energy forces the magnon to lie in a state within an energy window of width $\sim \Delta^{-1}$ measured from the center of the magnon band. This phase space restriction forces the magnon creation/decay rate to scale as $1/\Delta$, and (by the crossover time analysis above) implies that $D(\gamma, \Delta) \alt 1/\sqrt{\Delta}$.

While our analysis was phrased in terms of a particular scattering process (which we expect to be the leading one), it is clear that \emph{any} scattering process creating a magnon out of higher strings will acquire the same bandwidth restriction, so this bound applies to all local Hamiltonian perturbations.
For the specific subclass of nearest-neighbor or on-site perturbations one can derive a stronger bound that combines the two arguments above. For these, the matrix element for tunneling to a configuration with broken magnon number is itself suppressed by $1/\Delta$, as we discussed above for on-site noise. Combining this with the phase space restriction, we find that the crossover timescale grows at least as $\Delta^3$, giving the bound $D(\Delta) \alt 1/\Delta^{3/2}$. 

To test these predictions we have simulated the spin chains given by the Hamiltonian $H= H_{\rm XXZ}  + V$ where the XXZ Hamiltonian is either perturbed with integrability-breaking staggered couplings $V= \delta J \sum_i  (-1)^i (S^x_i S^x_{i+1} + S^y_i S^y_{i+1} + \Delta S^z_i S^z_{i+1}) $ or next-nearest-neighbor couplings $V= J' \sum_i (S^x_i S^x_{i+2} + S^y_i S^y_{i+2} + \Delta S^z_i S^z_{i+2})$.
Again, we simulated the dynamics using MPO methods. For Hamiltonian perturbations, as opposed to noisy perturbations, the simulation complexity (captured by the bond dimension of the MPO) grows exponentially in time. Therefore, our simulations are limited to relatively early times and cannot extract the saturated diffusion constant; nevertheless, they clearly display the non-monotonicity of $D(t)$ and the onset of the subdiffusive regime, where $D(t) \sim \kappa/\sqrt{t}$ (Fig.~\ref{hamfig}), with $\kappa$ approximately independent of $\Delta$.

\emph{Discussion}.---In this work we have presented evidence that integrability-breaking has drastic effects on transport in integrable systems where the integrable limit is itself diffusive, in sharp constraint with the ballistic case, as for the case of free fermions, see~\cite{suppmat}. For infinitesimal integrability breaking parameter $\gamma$, the diffusion constant jumps to a value that is independent of $\gamma$, but is parametrically lower than the integrable diffusion constant at large $\Delta$. The mechanism for this abrupt change in the diffusion constant is the emergence of a large subdiffusive temporal regime, which becomes the asymptotic behavior in the limit of large anisotropy. We presented a kinetic argument for this asymptotic subdiffusive behavior, in terms of the diffusive propagation of magnons whose magnetization is screened by thermal fluctuations. 

 We expect such non-monotonicity of the diffusion constant and the $\sqrt{\omega}$ dependence of the conductivity at low-frequencies to be observable in cold atomic settings with emergent XXZ interaction, see~\cite{Jepsen:2020aa,2105.01597,2021arXiv210804845B,2107.14459}, and in anisotropic Heisenberg-Ising compounds~\cite{PhysRevLett.123.067202,Rams2020}.

Our findings also clarify the reasons behind the apparent difficulties encountered in evaluating diffusion constants by dissipative truncation schemes, as introduced in \cite{2004.05177}. One should indeed expect that close to integrability, the diffusion constant is not a continuous function of the dissipation strength, making it hard to extrapolate its value in the limit of small noise. Our results could also be related to the vanishing of the diffusion constant $D\sim \Delta^{-1}$ obtained coupling an XXZ chain to boundary Lindblad spin reservoirs~\cite{PhysRevLett.106.220601}.

\acknowledgements{{\it Acknowledgements.}---We thank Vedika Khemani and Marcos Rigol for helpful discussions, and Aaron Friedman and Hansveer Singh for collaborations on related topics.  This work was supported by the National Science Foundation under NSF Grant No. DMR-1653271 (S.G.), the Air Force Office of Scientific Research under Grant No. FA9550-21-1-0123 (R.V.), and the Alfred P. Sloan Foundation through a Sloan Research Fellowship (R.V.). This research was performed while B.W. held an NRC postdoctoral fellowship at the National Institute of Standards and Technology. Some of the numerical MPO results were obtained via the ITensor library \cite{itensor}. }

\bibliography{SSD,refs}

\onecolumngrid
\newpage


\section*{Supplementary material}

# Supplemental material for "Subdiffusive hydrodynamics of nearly integrable anisotropic spin chains"

Jacopo De Nardis,[1] Sarang Gopalakrishnan,[2] Romain Vasseur,[3] and Brayden Ware[4, 5]

[1]*Laboratoire de Physique Théorique et Modélisation, CNRS UMR 8089, CY Cergy Paris Université, 95302 Cergy-Pontoise Cedex, France*
[2]*Department of Physics, The Pennsylvania State University, University Park, PA 16802, USA*
[3]*Department of Physics, University of Massachusetts, Amherst, MA 01003, USA*
[4]*Joint Quantum Institute, NIST/University of Maryland, College Park, Maryland 20742, USA*
[5]*Joint Center for Quantum Information and Computer Science, NIST/University of Maryland, College Park, Maryland 20742, USA*
(Dated: October 15, 2021)

## I. STOCHASTIC KINEMATICALLY CONSTRAINED MODEL WITH DOMAIN WALL NUMBER CONSERVATION

In this section, we consider a stochastic kinetically constrained model, where the total number of domain walls is strictly conserved, and the only allowed spin moves are those that respect this conservation law in addition to charge conservation. The discrete time evolution can be constructed from a brick-wall pattern of non-unitary gates that implement all possible transitions allowed by symmetries with equal probability.

For interactions acting on $M = 4$ sites, the allowed moves are flip-flop spin flips between sites $j$ and $j+1$ if the spins on sites $j-1$ and $j+2$ are aligned. Similar constraints have been considered in Refs. 1–6, and this model was studied in detail in Ref. 7. The allowed moves constrain the motion of the domain walls themselves, so that isolated domain walls are frozen, but adjacent pairs of domain walls (corresponding to isolated magnons in some background) can hop. In that case, the number of magnons (and more generally the string number) is conserved by the evolution. This corresponds to a form of intermediate integrability breaking where the string number is conserved, but magnons move diffusively (while larger strings are frozen). This is the case relevant to integrability-breaking in the XXZ spin chain discussed in the main text, and this model was shown to have subdiffusive spin dynamics with $z = 4$ in Ref. 7.

However, subdiffusion *does not* follow from charge and domain wall conservation alone. To illustrate this point, we classically simulated the stochastic kinetically constrained models with all possible charge and domain wall conserving transitions within an interaction radius of $M = 4, 5$, and 6 sites. For each model, we computed the average spin structure factor by averaging over $7 \times 10^6$ realizations of the dynamics. As noted in the main text, for $M > 4$, the allowed spin moves do not conserve magnon number: for example, they connect the configurations $\ldots \downarrow\downarrow\uparrow\uparrow\uparrow\downarrow\downarrow\uparrow\downarrow \ldots$ and $\ldots \downarrow\downarrow\uparrow\uparrow\downarrow\downarrow\uparrow\uparrow\downarrow \ldots$, turning two 2-strings into a magnon and a 3-string and *vice versa* while conserving both charge

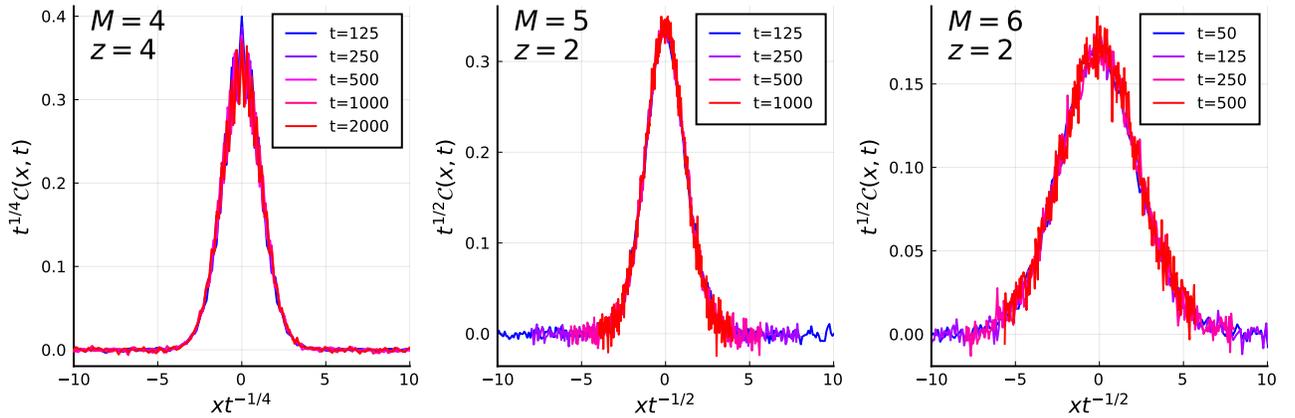

FIG. 1: **Classical kinetically constrained models.** Structure factor $\mathcal{C}(x,t) = \langle \sigma^z(x,t)\sigma^z(0,0) \rangle$ in a stochastic model with charge and domain wall conservation, with local interactions (non-unitary "gates") acting on $M$ sites. For $M = 4$, which is the case relevant to the XXZ spin chain with broken integrability, the magnon number is conserved and we find subdiffusive transport with $z = 4$ (left). For higher-range interactions, the magnon number is not conserved, and we find ordinary diffusion $z = 2$, indicating that domain wall conservation is *not* enough for subdiffusion to occur.

and domain wall number. This breaks integrability strongly, and in that case, our numerical results (Fig. 1) indicate vanilla diffusion with dynamical exponent $z = 2$.

## II. SPIN TRANSPORT IN THE NOISY XX CHAIN: BOLTZMANN APPROACH

In the main text, we focused on the fate of integrability breaking perturbations on spin transport in the non-trivial case where the integrable limit shows spin diffusion. In the more generic case where spin transport is ballistic in the integrable limit, the effect of integrability breaking perturbations is a lot simpler: the Drude weight is broadened into a generalized Lorentzian by the perturbation. To illustrate this, we will work out the case of the XX quantum spin chain subject to noise, and refer the reader to the recent references 8–11 for generalizations to interacting cases. We focus on XX spin chain, which can be mapped onto free fermions with dispersion relation $\epsilon_k = -\cos k$, subject to dephasing noise coupling to $S_i^z = c_i^\dagger c_i - 1/2$. We consider spatially correlated noise, characterized by a function $G(x)$, but uncorrelated in time. The Boltzmann equation reads

$$\partial_t n_k + v_k \partial_x n_k = I_k[n] = \gamma \int_{-\pi}^{\pi} \frac{dk'}{2\pi} \hat{G}(k-k')(n_{k'}(1-n_k) - n_k(1-n_{k'})), \tag{1}$$

where $\hat{G}$ is the Fourier transform of $G$. In the following, it will be convenient to introduce the decay matrix

$$\Gamma_{kk'} = -2\pi \frac{\delta I_k}{\delta n_{k'}} = -\gamma \hat{G}(k-k') + \delta(k-k')\gamma \int_{-\pi}^{\pi} dq \hat{G}(q) = -\gamma \left( \hat{G} - \hat{\delta} \int \hat{G} \right)_{k,k'}. \tag{2}$$

This is a Markov matrix, and we have $\int \frac{dk'}{2\pi} \Gamma_{kk'} = 0$, corresponding to an eigenvalue 0 (zero mode) for this matrix which is associated with particle number conservation. The spin conductivity can then be computed using the Kubo formula

$$\sigma = \beta \int_0^\infty dt \int \frac{dk}{2\pi} \frac{dk'}{2\pi} v_k v_{k'} \int dx \langle \delta n_k(x,t) \delta n_{k'}(0,0) \rangle. \tag{3}$$

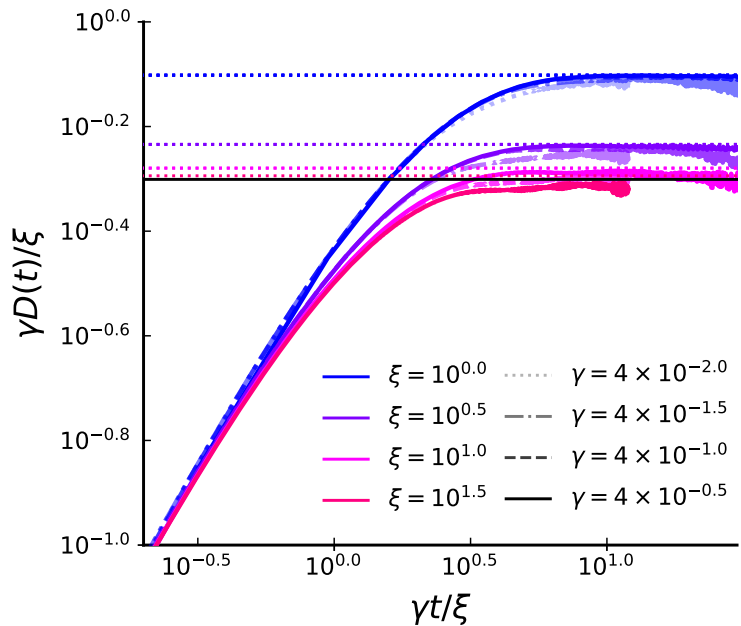

FIG. 2: **Time-dependent diffusion in the noisy XX spin chain.** We plot the diffusion constant $D(t)$ for evolution in the presence of correlated dephasing noise of strength $\gamma$ and with correlation length $\xi$ for a number of $\gamma$ and $\xi$. The saturated value of $\gamma D(t)/\xi$ from numerically integrating Eq. 4 depends only on $\xi$ and is shown by the dotted lines. (These values approach $1/2$ as $\xi \to \infty$.) In all cases, the TEBD data shows saturation to the predicted values.

where $v_k = \sin k$, $\langle \delta n_k(x,t) \delta n_{k'}(0,0) \rangle = \int \frac{dk''}{2\pi} \left[ e^{-\hat{\Gamma} t} \right]_{kk''} \int dx \langle \delta n_{k''}(x,0) \delta n_{k'}(0,0) \rangle$, with the initial (thermal) Fermi factor fluctuations: $\int dx \langle \delta n_{k''}(x,0) \delta n_{k'}(0,0) \rangle = 2\pi n_{k'}(1 - n_{k'}) \delta(k' - k'')$. This yields

$$\sigma = \frac{\beta}{4} \int_{-\pi}^{\pi} \frac{dkdk'}{(2\pi)^2} \sin k \sin k' \Gamma^{-1}_{kk'}, \qquad (4)$$

where we have used $n_k = 1/2$ at infinite temperature. The conductivity is related to the spin diffusion constant $D$ by the Einstein relation $\sigma = \chi D$ with $\chi = \beta/4$ the spin susceptibility.

We consider noise spatially correlated over a length $\xi$, with $G(x) = e^{-|x|/\xi}$ so that $\hat{G}(k) = \frac{2\xi}{1+\xi^2 k^2}$. To compute the diffusion constant, we need the (pseudo)inverse $\left( 2\pi \gamma \hat{\delta} - \gamma \hat{G} \right)^{-1}_{k,k'}$ where we have used $\int \hat{G} = 2\pi$, where the inverse is defined by $\int \frac{dk'}{2\pi} \Gamma_{k,k'} \Gamma^{-1}_{k',k''} = 2\pi \delta(k-k'')$ with $\hat{\Gamma} = \gamma \left( 2\pi \hat{\delta} - \hat{G} \right)$. This inverse can be computed by going to real space $\Gamma^{-1}(x) = \frac{1}{\gamma(1-G(x))}$. Going back to $k$-space and considering the limit of smooth noise ($\xi \gg 1$) yields

$$\Gamma^{-1}(k) \sim \frac{\xi}{\gamma} \sum_{n=1}^{\infty} \frac{2 \cos kn}{n} = -\frac{\xi}{\gamma} \log\left[2(1-\cos k)\right]. \qquad (5)$$

In this limit, the integral in Eq. 4 can be computed exactly, resulting in a diffusion constant $D = \frac{\xi}{2\gamma}$. For general $\xi$, the integral can be done numerically. As shown in Fig. 2, the resulting diffusion constant exactly matches the saturated value of $D(t)$ computed from the variance of the structure factor in TEBD calculations. In contrast with the case studied in the main text, the diffusion constant depends on the noise strength $\gamma$, and has a simple interpretation as the broadened Drude weight of the integrable case.

## III. NUMERICAL DATA FOR THE LOW-FREQUENCY CONDUCTIVITY $\sigma(\omega)$

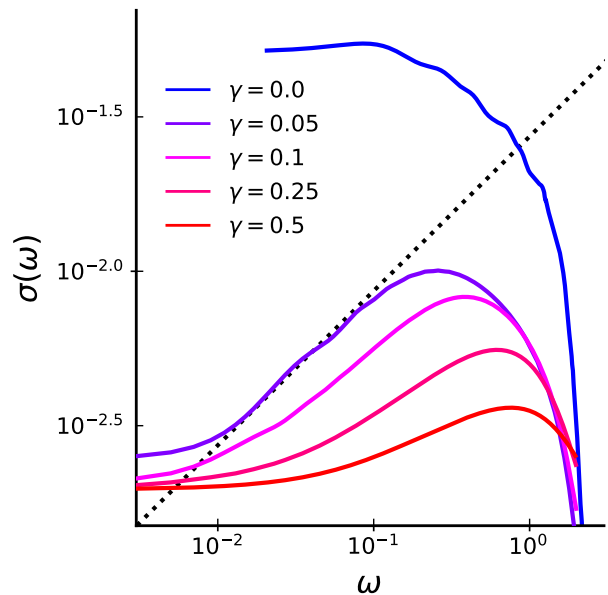

FIG. 3: **Low-frequency spin conductivity in the noisy XXZ chain.** Plot of $\sigma(\omega)$ obtained from MPO simulations at $\Delta = 8$ and for various values of the noise strengths $\gamma$. The dashed line corresponds to the predicted intermediate subdiffusive behavior $\sim \omega^{1/2}$. Note that spectral weights is also redistributed by the noise at higher frequencies, not shown in this plot.

In the main text, we presented numerical plots of the time-dependent diffusion constant $D(t)$, obtained from the spatial variance of the spin structure factor. For completeness, we also show in Fig. 3 the low-frequency conductivity

obtained numerically from the Kubo formula, for a representative value $\Delta = 8$.



\end{document}